\documentclass[preprint,aps,showpacs,amsfonts,epsf]{revtex4}

\input{epsf.tex}

\renewcommand{\a}{\alpha}

\newcommand{\nabv}{\mbox{\boldmath$\nabla$}}
\newcommand{\rhov}{\mbox{\boldmath$\rho$}}

\newcommand{\be}{\begin{equation}}
\newcommand{\ee}{\end{equation}}
\newcommand{\bea}{\begin{eqnarray}}
\newcommand{\eea}{\end{eqnarray}}
\newcommand{\ba}{\begin{array}}
\newcommand{\ea}{\end{array}}
\def\J#1#2#3#4{{#1} {\bf #2}, #3 (#4)}

\def\CQG{Class. Quantum Grav.}

\def\GRG{Gen. Relativ. Grav.}
\def\PLA{Phys. Lett. A}

\begin{document}
\draft
\title{On a static charged fluid around a magnetized mass}

\author{I.~Cabrera--Munguia and V.~S.~Manko}
\address{Departamento de F\'\i sica, Centro de Investigaci\'on y de
Estudios Avanzados del IPN, A.P. 14-740, 07000 M\'exico D.F.,
Mexico}

\begin{abstract}
We show that {\it any} magnetostatic axially symmetric solution of the Einstein--Maxwell equations can be endowed with a specific charged fluid source of the Polanco et al type via a simple procedure requiring the knowledge of exclusively the magnetostatic seed spacetime. Using this procedure we construct yet another exact solution for a massive magnetic dipole surrounded by a static charged fluid which is different from the Polanco et al metric. \end{abstract}


\maketitle


\section{Introduction}

In a recent paper \cite{PLU} Polanco et al have generalized the Gutsunaev--Manko massive magnetic dipole solution \cite{GMa} to the case when the dipole is surrounded by a charged static fluid. The new solution contains an additional parameter $\Phi_0$ defining the electric field, and its fluid source exhibits an interesting physical property -- it compensates the frame dragging caused by a charged magnetic dipole \cite{Bon}, so that the metric remains static in spite of the presence of both the electric and magnetic fields. Since some of the field equations have been solved in \cite{PLU} only for the particular case determined by the Gutsunaev--Manko two--parameter metric, it is the aim of our paper to demonstrate that in principal {\it any} static axisymmetric magnetovac solution of the Einstein--Maxwell equations can be generalized in the Polanco et al manner with the aid of the following simple procedure:

Given an arbitrary magnetostatic solution of the Einstein--Maxwell equations defined by the functions $f$, $\gamma$ and $A_\phi$, where $f$ and $\gamma$ are metric coefficients in the Weyl line element \be d s^2=f^{-1}[e^{2\gamma}(d\rho^2+d z^2) +\rho^2 d\varphi^2]-f d t^2, \label{W1} \ee and $A_\phi$ the magnetostatic potential, its ``static charged fluid generalization'' of the Polanco et al type is defined by the functions $\tilde f$, $\tilde\gamma$, $\tilde A_\phi$ and $\tilde A_t$ obtainable from $f$, $\gamma$ and $A_\phi$ with the aid of the formulas \be \tilde f=f, \quad \tilde\gamma=(1-\Phi_0^2)\gamma, \quad \tilde A_\phi=(1-\Phi_0^2)^{1/2}A_\phi, \quad \tilde A_t=\Phi_0 f^{1/2}, \quad \Phi_0={\rm const}, \label{proc} \ee where $\tilde f$ and $\tilde\gamma$ are metric coefficients in the line element \be d s^2=\tilde f^{-1}[e^{2\tilde\gamma}(d\rho^2+d z^2) +\rho^2 d\varphi^2]-\tilde f d t^2, \label{W2} \ee and $\tilde A_\phi$, $\tilde A_t$ are, respectively, the magnetic and electric components of the electromagnetic 4--potential $A_i=(0,0,\tilde A_\phi,\tilde A_t)$.

In the next section we will prove the validity of the formulas (\ref{proc}) by directly comparing the field equations of the paper \cite{PLU} with those defining the magnetostatic solutions in the axially symmetric case, and in Sec.~III we apply the solution--generating procedure (\ref{proc}) to the magnetic dipole solution \cite{GMa2} whose potential $A_\phi$ is obtained by us for the first time.

\section{Reduction of the Polanco et al static fluid problem to the system of magnetostatic equations}

In order to understand better the structure of the field equations derived in the paper \cite{PLU}, let us first consider the magnetostatic equations. The magnetovac static spacetimes with axial symmetry are defined by the line element (\ref{W1}) and by the following set of the Einstein--Maxwell equations (see, e.g., \cite{GMa2}) \bea f\Delta f&=&(\nabv f)^2+2\rho^{-2}f^3(\nabv A_\phi)^2, \label{M1} \\ f\Delta A_\phi&=&2\rho^{-1}f  A_{\phi,\rho} -\nabv f\nabv A_\phi, \label{M2} \\ 4\gamma_{,\rho}&=&\rho f^{-2}(f_{,\rho}^2-f_{,z}^2)+4\rho^{-1}f(A_{\phi,\rho}^2- A_{\phi,z}^2), \label{M3} \\ 2\gamma_{,z}&=&\rho f^{-2}f_{,\rho} f_{,z}+4\rho^{-1}f A_{\phi,\rho} A_{\phi,z}, \label{M4} \eea where $\Delta\equiv\partial_{\rho\rho}+\rho^{-1}\partial_{\rho}+\partial_{zz}$ and $\nabv\equiv\rhov_0\partial_{\rho}+{\bold z}_0\partial_{z}$ ($\rhov_0$ and ${\bold z}_0$ are unit vectors). A particular set of the functions $f$, $\gamma$, $A_\phi$ satisfying Eqs.~(\ref{M1})--(\ref{M4}) fully determines the corresponding magnetostatic spacetime.

Let us turn now to the field equations of the paper \cite{PLU}. In what follows, we will change the Polanco et al's functions $f$, $\Lambda$, $\Phi$ and $W$ to, respectively, ours $\tilde f$, $\tilde\gamma$, $\tilde A_t$ and $\tilde A_\phi$; we also note that the ``divergence'' operation is defined in \cite{PLU} in a slightly different way than usually adopted for the cylindrical coordinates, so the main equations of \cite{PLU} will be rewritten in the same fashion as Eqs.~(\ref{M1})--(\ref{M4}) above, i.e., avoiding the ``divergence'' operation. After these preliminary remarks and taking into account that Eq.~(25) of Ref.~\cite{PLU} has already been shown by Polanco et al to be satisfied by the function $\Phi=\Phi_0 f^{1/2}$ which in our notations rewrites as \be \tilde A_t=\Phi_0\tilde f^{1/2}, \label{At} \ee the field equations for the remaining functions $\tilde f$, $\tilde A_\phi$ and $\tilde\gamma$ take the form \bea (1-\Phi_0^2)\tilde f\Delta\tilde f&=&(1-\Phi_0^2)(\nabv\tilde f)^2+2\rho^{-2}\tilde f^3(\nabv\tilde A_\phi)^2, \label{P1} \\ \tilde f\Delta\tilde A_\phi&=&2\rho^{-1}\tilde f \tilde A_{\phi,\rho} -\nabv\tilde f\nabv\tilde A_\phi, \label{P2} \\ 4\tilde\gamma_{,\rho}&=&(1-\Phi_0^2)\rho \tilde f^{-2}(\tilde f_{,\rho}^2-\tilde f_{,z}^2)+4\rho^{-1}\tilde f(\tilde A_{\phi,\rho}^2-\tilde A_{\phi,z}^2), \label{P3} \\ 2\tilde\gamma_{,z}&=&(1-\Phi_0^2)\rho \tilde f^{-2}\tilde f_{,\rho}\tilde f_{,z}+4\rho^{-1}\tilde f\tilde A_{\phi,\rho}\tilde A_{\phi,z}. \label{P4} \eea

It is easy to see now that the substitutions \be \tilde f=f, \quad \tilde\gamma=(1-\Phi_0^2)\gamma, \quad \tilde A_\phi=(1-\Phi_0^2)^{1/2}A_\phi \label{sub} \ee cast Eqs.~(\ref{P1})--(\ref{P4}) into the magnetostatic system (\ref{M1})--(\ref{M4}). Therefore, any solution $f$, $\gamma$ and $A_\phi$ of the latter system immediately yields the corresponding solution $\tilde f$, $\tilde\gamma$, $\tilde A_\phi$ and $\tilde A_t$ of Eqs.~(\ref{P1})--(\ref{P4}) through the formulas (\ref{proc}).

In some sense, the solution--generating procedure (\ref{proc}) is a trivial one because it does not require any technical work for the formal introduction of specific charged fluid sources into the known magnetostatic solutions. The example which will be considered in the next section is to a certain extent an exception since it involves a preliminary calculation of the potential $A_\phi$ for the seed magnetovac solution.

\section{A non--trivial example}

Let us illustrate the use of the procedure (\ref{proc}) by taking as a seed magnetostatic spacetime the asymptotically flat solution \cite{GMa2}. The latter solution describes the field of a massive magnetic dipole and possesses the Schwarzschild pure vacuum limit in the absence of the magnetic field. It has a bit more complicated form than the other Gutsunaev--Manko solution \cite{GMa} utilized by Polanco et al in \cite{PLU}, however, it is also equatorially symmetric \cite{EMR}, unlike for instance two analogous solutions \cite{Man} for a magnetized mass. In the paper \cite{GMa2} the calculation of the potential $A_\phi$ was not carried out, the magnetic field being defined by the function $A'_\phi$ \cite{GMa3}. Hence, before applying the procedure (\ref{proc}), we have yet to find the form of the potential $A_\phi$ for the seed solution \cite{GMa2}.

In prolate spheroidal coordinates ($x$, $y$) related to the coordinates ($\rho$, $z$) via the formulas \be \rho=k\sqrt{(x^2-1)(1-y^2)}, \quad z=kxy \label{xy} \ee ($k$ is a real constant), the functions $f$, $\gamma$ and $A'_\phi$ of the seed metric have the form \cite{GMa2} \bea f&=&\frac{x-1}{x+1}\cdot\frac{A^2}{B^2}, \quad e^{2\gamma}=\frac{x^2-1}{x^2-y^2}\cdot\frac{A^2}{(1+\a^2)^8(x^2-y^2)^{24}}, \nonumber\\ A'_\phi&=&-8\a xy(x-1)[(x^2-y^2)^4+2\a^2(x^2+y^2)(x^2-1)^3]B^{-1}, \nonumber\\ A&=&[(x^2-y^2)^3+\a^2(x^2-1)^3]^2+4\a^2x^2(1-y^2)(x^2-1)^2(x^2+3y^2)^2, \nonumber\\ B&=&[(x^2-y^2)^3+\a^2(x+1)^2(x^2-1)^2]^2 \nonumber\\ &-&4\a^2y^2(x+1)^2(x^2-1)(x^3-3x^2+3xy^2-y^2)^2. \label{GM2} \eea

The corresponding expression of the potential $A_\phi$ can be found by integrating the system \cite{GMa3} \be A_{\phi,x}=k(y^2-1)f^{-1}A'_{\phi,y}, \quad A_{\phi,y}=k(x^2-1)f^{-1}A'_{\phi,x}, \ee and the final result for $A_\phi$ can be presented in the following two forms: \bea A_\phi&=&-\frac{4\a k(1-y^2)C}{(1+\a^2)A}, \nonumber\\ C&=&[(x^2-y^2)^3+\a^2(x^2-1)^3]\{(x^2-y^2)(2x^3+x^2+y^2) \nonumber\\ &+&\a^2(x+1)^2[2x(x^2-y^2-1)+y^2+1]\} \nonumber\\ &+&2\a^2x^2(x^2-1)^2(x^2+3y^2)[(1+\a^2)(x+1)^3-2x^2-6y^2], \label{A3_1} \eea which emphasizes the property of this $A_\phi$ to take zero values on the upper and lower parts of the symmetry axis ($y=\pm1$); and \bea A_\phi&=&-\frac{4\a k(x+1)^2D}{A}+\frac{4\a k}{1+\a^2}, \nonumber\\ D&=&(x^2-y^2)(x^2+y^2-2xy^2)[(x^2-y^2)^3+\a^2(x^2-1)^3] \nonumber\\ &+&2\a^2x^2(x-1)^2(x+1)^3(1-y^2)(x^2+3y^2), \label{A3_2} \eea which obviously is a more concise expression than (\ref{A3_1}), though at the expense of hiding the important factor ($1-y^2$).

Writing out the ``charged static fluid generalization'' of the solution (\ref{GM2}) with the aid of (\ref{proc}) is now immediate: \bea \tilde f&=&\frac{x-1}{x+1}\cdot\frac{A^2}{B^2}, \quad e^{2\tilde\gamma}=\left[\frac{(x^2-1)A^2}{(1+\a^2)^8(x^2-y^2)^{25}}\right]^{1-\Phi_0^2}, \nonumber\\ \tilde A_t&=&\Phi_0\left(\frac{x-1}{x+1}\right)^{1/2}\frac{A}{B}, \quad A_\phi=-\frac{4\a k(1-\Phi_0^2)^{1/2}(1-y^2)C}{(1+\a^2)A}, \label{Gen} \eea where the polynomials $A$, $B$ and $C$ have already been defined in (\ref{GM2}) and (\ref{A3_1}).

Solution (\ref{Gen}) exhibits essentially the same physical properties as the one considered by Polanco et al, so we refer the reader to the paper \cite{PLU} for further details.

\section*{Acknowledgments}

This work was supported by Project 45946--F from Consejo Nacional de Ciencia y Tecnolog\'\i a, Mexico.

\newpage

\end{document}